\documentclass{moriond}

\def\Journal#1#2#3#4{{#1} {\bf #2}, #3 (#4)}

\def\AA{\em A\&A}
\def\Ann{\em Annales d'Astrophysique}
\def\APh{\em APh}
\def\ApJ{\em ApJ}
\def\ApJL{\em ApJ Lett.}
\def\ApSS{\em Ap\&SS}
\def\CPC{\em Comp. Phys. Comm.}
\def\ExpA{\em Experimental Astronomy}
\def\JCAP{\em JCAP}
\def\JETP{\em  JETP}
\def\SJETP{\em Sov. Phys. JETP}
\def\JETPL{\em JETP Lett.}
\def\JPhysA{\em J. Phys. A: Gen. Phys.}
\def\JPCS{\em J. Phys. Conf. Ser.}
\def\MNRAS{\em MNRAS}
\def\Nat{\em Nature}
\def\NatP{\em Nature Physics}

\def\PLB{{\em Phys. Lett.}  B}
\def\PRL{\em Phys. Rev. Lett.}
\def\PRD{{\em Phys. Rev.} D}
\def\PR{\em Phys. Rev.}
\def\PTP{\em Prog. Theor. Phys.}
\def\Sci{\em Science}

\def\be{\begin{equation}}
\def\ee{\end{equation}}
\def\bea{\begin{eqnarray}}
\def\eea{\end{eqnarray}}

\newcommand{\Photo}{}

\begin{document}
\vspace*{4cm}
\title{THE INTERGALACTIC ELECTROMAGNETIC CASCADE SOLUTION \\ FOR THE ANOMALIES FROM $\gamma$-RAY BLAZAR OBSERVATIONS}

\author{T.A. DZHATDOEV}

\address{Federal State Budget Educational Institution of Higher Education \\ M.V. Lomonosov Moscow State University, Skobeltsyn Institute
of Nuclear Physics (SINP MSU), \\ 1(2), Leninskie gory, GSP-1, Moscow 119991, Russian Federation and \\
Institute for Cosmic Ray Research, University of Tokyo, Tokyo, Japan}
%∗e-mail: timur1606@gmail.com
\maketitle
\abstracts{Recent progress in very high energy (VHE, $E>$100 $GeV$) $\gamma$-ray observations, together with advances in the extragalactic background light (EBL) modelling, allows to search for new phenomena such as $\gamma$-axion-like particle ($\gamma \rightarrow ALP$) oscillation and to explore the extragalactic magnetic field (EGMF) strength and structure. These studies are usually performed by searching for some deviation from the so-called absorption-only model, that accounts for only primary photon absorption on the EBL and adiabatic losses. In fact, there exist se\-ve\-ral indications that the absorption-only model is incomplete. We present and discuss the intergalactic electromagnetic cascade model (IECM) --- the simplest model that allows to coherently explain all known anomalies. This model has a number of robust signatures that could be searched for with present and future instruments. The IECM model may serve as a new background template, allowing to make future searches for $\gamma \rightarrow ALP$ oscillation more robust. A detailed account of our calculations is available in astro-ph/1609.01013v2 (A\&A, in print).}

\section{Introduction}

Blazars are the brightest distant (redshift $z_{0}>$0.03) extragalactic high energy (HE, $E>$100 MeV) \cite{r01}--\cite{r02} and very high energy (VHE, $E>$100 $GeV$) \cite{r03} $\gamma$-ray emitters. Observable spectra of these sources are sensitive to properties of the intervening extragalactic background light (EBL) \cite{r04}--\cite{r07} and extragalactic magnetic field (EGMF) \cite{r08}--\cite{r10}. Primary $\gamma$-rays with an energy $E_{0}>$1 TeV and $z_{0}>$0.1 are strongly absorbed on EBL photons (e.g. \cite{r11}--\cite{r15}); secondary electrons and positrons (hereafter simply ``electrons'') produce secondary (cascade) photons through the inverse Compton (IC) process.

In this report a brief discussion of extragalactic $\gamma$-ray propagation models is presented with emphasis on possible effects imprinted to observable spectra of blazars by the development of electromagnetic (EM) cascades in the intergalactic volume. The author readily acknowledges that, due to limited space available, this overview is by no means exhaustive or unbiased.

\section{Extragalactic $\gamma$-ray propagation models} \label{sec:models}

Most of existing models that describe the transformation of $\gamma$-ray spectrum during extragalactic propagation may be divided to the following three  classes: \\
1) the absorption-only model which includes only the $\gamma\gamma$ pair production (PP) process and adiabatic losses of primary $\gamma$-rays \\
2) intergalactic cascade models, namely: \\
2a) the electromagnetic (EM) cascade model which accounts for the PP and IC processes, as well as adiabatic losses, assuming that primary particles are $\gamma$-rays or electrons \\
2b) the hadronic cascade model which accounts for the PP and IC processes, as well as adiabatic losses, assuming that primary particles are protons or nuclei of ultra-high energy (UHE, $E_{0}>$1 $EeV$) that could produce $\gamma$-rays and electrons via the photohadronic and the Bethe-Heitler pair production processes with subsequent development of EM cascades \\
3) exotic models which postulate some new physics, dramatically changing the mode of extragalactic $\gamma$-ray propagation, namely: \\
3a) the gamma-axion-like particle ($\gamma \rightarrow ALP$) oscillation process \\
3b) Lorentz invariance violation (LIV) \\
3c) exotic primaries or any other non-conventional effects imaginable.

The absorption-only model was historically the first one; it is currently the most well-established and by far the most commonly used extragalactic $\gamma$-ray propagation model. Not long after the first work on the astrophysical implications of the $\gamma\gamma$ PP process \cite{r04}, it was realised that UHE $\gamma$-rays may absorb on the universal radio background (URB) photons \cite{r16}. Almost immediately after the discovery of the cosmic microwave background (CMB) it was understood that this dense (compared to the EBL) photon field constitutes a target for photons with energy $E>$100 $TeV$ \cite{r05}--\cite{r06}. Soon after the discovery of the first TeV $\gamma$-ray emitting blazar \cite{r17}, the first $\gamma$-astronomical constraints on the EBL number density were obtained \cite{r18}--\cite{r19}, assuming the absorption-only model. Further constaints using this method include \cite{r20}--\cite{r22}.

However, there exist some deviations from the absorption-only model, hereafter referred to as the ``anomalies'', even if their explanation does not call for any new physics \cite{r23}--\cite{r26}. These effects are still not very well established. Namely, the statistical significance of the high-energy anomaly \cite{r23} may strongly and non-trivially depend on the assumed EBL spectral shape and intensity normalization \cite{r24}; the significance of the anomalies found in \cite{r25}--\cite{r26} is modest, at the level of 2-3 $\sigma$. Below we argue that all these effects find their natural explanation in the framework of the intergalactic electromagnetic cascade model (IECM). As well as the absorption-only model, the  IECM has a long history. As early as in 1966, it was already clearly understood that intergalactic EM cascades may contribute to observable $\gamma$-ray emission of point-like sources \cite{r06}. Various aspects of the IECM were investigated in numerous works, including \cite{r27}-\cite{r28},\cite{r08},\cite{r29},\cite{r09}-\cite{r10},\cite{r30}-\cite{r38}. Up to the author's knowledge, the work \cite{r39} was the first where the intergalactic hadronic cascade model was applied to the highest-energy region of blazar spectra; from 2010 on, many such papers were published, including \cite{r40},\cite{r35}--\cite{r36},\cite{r41},\cite{r38}.

Concerning the $\gamma \rightarrow ALP$ model, let us mention only the latest works \cite{r42}--\cite{r43}, as well as the detailed treatise \cite{r44}; LIV effects were considered, among others, in \cite{r45}--\cite{r46}. Many ``exotic'' models may be excluded or, at least, may have their parameters strongly constrained. For instance, the hypothesis that showers with multi-TeV primary energy observed with the HEGRA Cherenkov telescope array are pure Bose-Einstein condensates (BECs) \cite{r47} was rejected in \cite{r48}. BEC, which represents a superposition of several or many photons, usually develop a shower in the atmosphere earlier than individual $\gamma$-rays; this would affects the parameters of the angular images observed by the HEGRA detector. Such a change of the distributions of the parameters was not observed, and, therefore, the primary BEC hypothesis was experimentally ruled out.

In what follows, synchrotron energy losses of cascade electrons in voids of the large scale structure (LSS) are neglected, as the primary $\gamma$-ray energy is typically below 1 $PeV$ (see Fig. 7 of \cite{r37}). The role of collective effects in $e^{+}e^{-}$ intergalactic EM cascade beams is still not well established, notwithstanding the 60-year age of the subject \cite{r49}--\cite{r50}. More recently the interest in this area was raised by \cite{r51}; recent works on the subject include \cite{r52}--\cite{r56}; some other references could be found in \cite{r38}, subsection 2.1. Such effects are neglected for the rest of this paper; calculations presented below are for $z$= 0.186 and the EGMF strength $B$=0, unless otherwise stated. The effects induced by non-zero EGMF are discussed in sect.~\ref{sec:flood}.

\section{Electromagnetic cascade in the expanding Universe} \label{sec:cascade}

\begin{figure}
\begin{minipage}{0.90\linewidth}
\centerline{\includegraphics[width=0.9\linewidth]{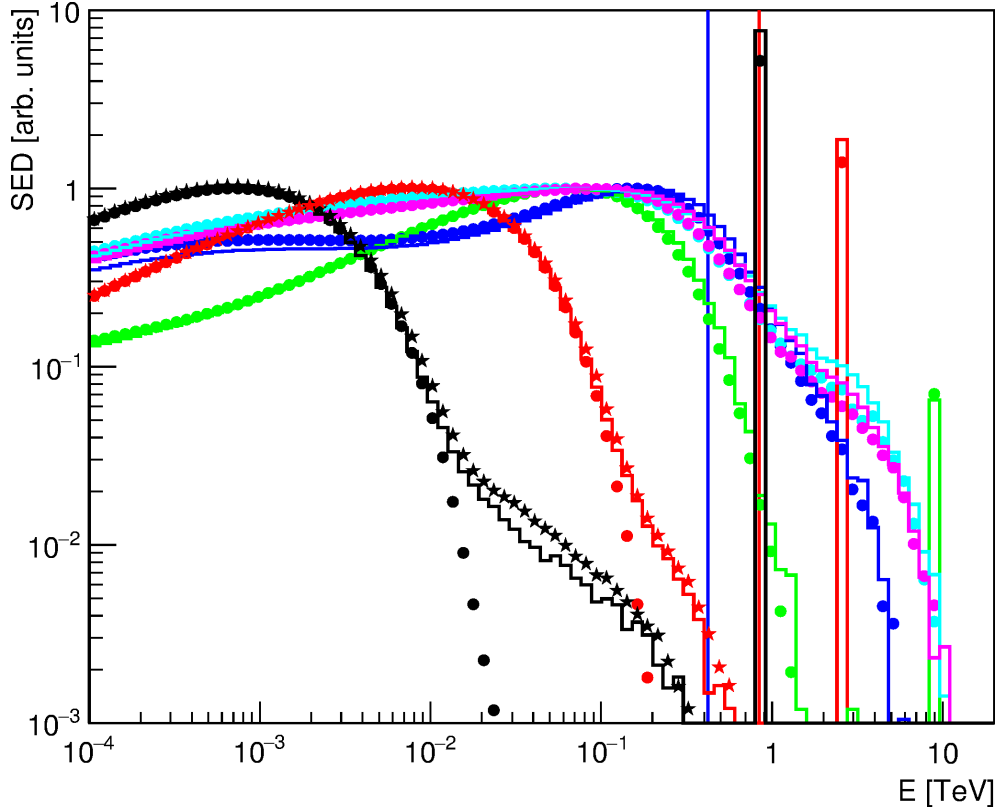}}
\end{minipage}
\caption[]{Observable angle-averaged spectra produced by primary monoenergetic $\gamma$-rays: $E_{0}$= 1 $TeV$ (black), 3 $TeV$ (red), 10 $TeV$ (green), 30 $TeV$ (blue), 100 $TeV$ (cyan), 1 $PeV$ (magenta). Histograms --- results obtained with the ELMAG 2.02 code, circles --- with the ECS 1.0 code (IC on the CMB only), stars (for $E_{0}$= 1 $TeV$ and 3 $TeV$)--- with the ECS 1.0 code (IC on both CMB and EBL). Vertical red line shows the value of the primary energy redshifted to the observer's frame (1 $TeV/(1+z)$), blue line --- half this value.}
\label{fig:delta}
\end{figure}

For the case of the primary energy (which is hereafter defined in the source rest-frame) $E_{0}<$1 $EeV$ and z$\approx$ 0.2 there are two distinct regimes of intergalactic EM cascade development: the ``one-generation regime'' for $E_{0}<$10 $TeV$ and the ``universal regime'' for $E_{0}>$100 $TeV$ \cite{r37}--\cite{r38}. Both regimes are clearly identifiable in Fig.~\ref{fig:delta}.  Calculations were performed with the ELMAG 2.02 publicly-available code \cite{r57}, assuming the EBL model of \cite{r13} (see \cite{r38}, discussion of Fig. 1), and the new code ECS 1.0 (from ``electromagnetic cascade spectrum'') developed by the author \cite{r58} with the EBL model of \cite{r15}. Normalization is to the maximum of the cascade component in each case. The overall agreement between the results obtained with the two different codes is reasonable. For the case of $E_{0}$=1 $TeV$ and 3 $TeV$ an additional component of relatively high-energy cascade photons produced on the EBL is clearly seen. Even this component, however, is almost extinguished above the energy $E_{0}/(2(1+z))$. A more detailed discussion of relevant physics may be found in \cite{r38}.

\section{The signatures of the intergalactic electromagnetic cascade model} \label{sec:signatures}

\begin{figure}
\begin{minipage}{0.90\linewidth}
\centerline{\includegraphics[width=0.7\linewidth]{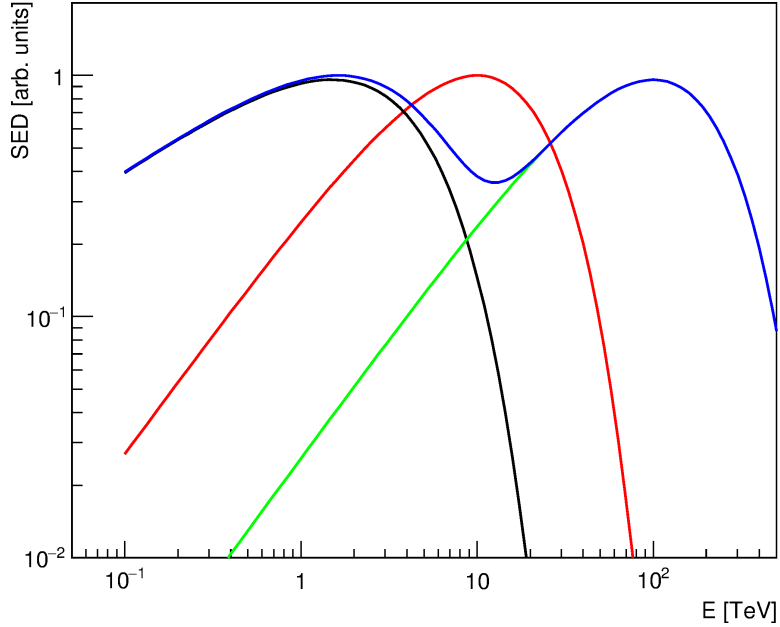}}
\end{minipage}
\caption[]{Schematic representation of two primary $\gamma$-ray spectrum scenarious considered in this study: red line --- {\it scenario 1}, blue line --- {\it scenario 2}. For the case of the {\it scenario 2} there are two components: the relatively low-energy leptonic component (black line) and the high-energy hadronic component (green line).}
\label{fig:spectra}
\end{figure}

Our study is mostly devoted to extreme $TeV$ blazars \cite{r59}. Below we consider two different sce\-na\-ri\-ous of their intrinsic (primary) $\gamma$-ray spectra, shown in Fig.~\ref{fig:spectra}: 1) hard spectrum up to the energy $\sim$10 $TeV$, allowing a substantial cascade contribution at energy $\sim$100 $GeV$ to the observable spectrum 2)~a two-component spectrum with a pile-up around $\sim$100 $TeV$, so that the cascades develop in the universal regime. The {\it scenario 2} is motivated by the recent findings indicating that some blazars may be responsible for a part of Ice Cube astrophysical neutrinos \cite{r60}--\cite{r61}. In this case neutrino production is accompanied by associated intrinsic $\gamma$-rays of similar energy \cite{r28}. A recent study \cite{r62} shows that hadronic processes may cause a hardeding in the intrinsic spectrum of an extreme $TeV$ blazar. A modest pile-up of hadronic nature may appear even in the spectra of some ``classical'' blazars, such as {\it Mkn 421}, {\it Mkn 501}, or {\it PKS 2155-304} \cite{r63}.

\begin{figure}
\begin{minipage}{0.50\linewidth}
\centerline{\includegraphics[width=0.9\linewidth]{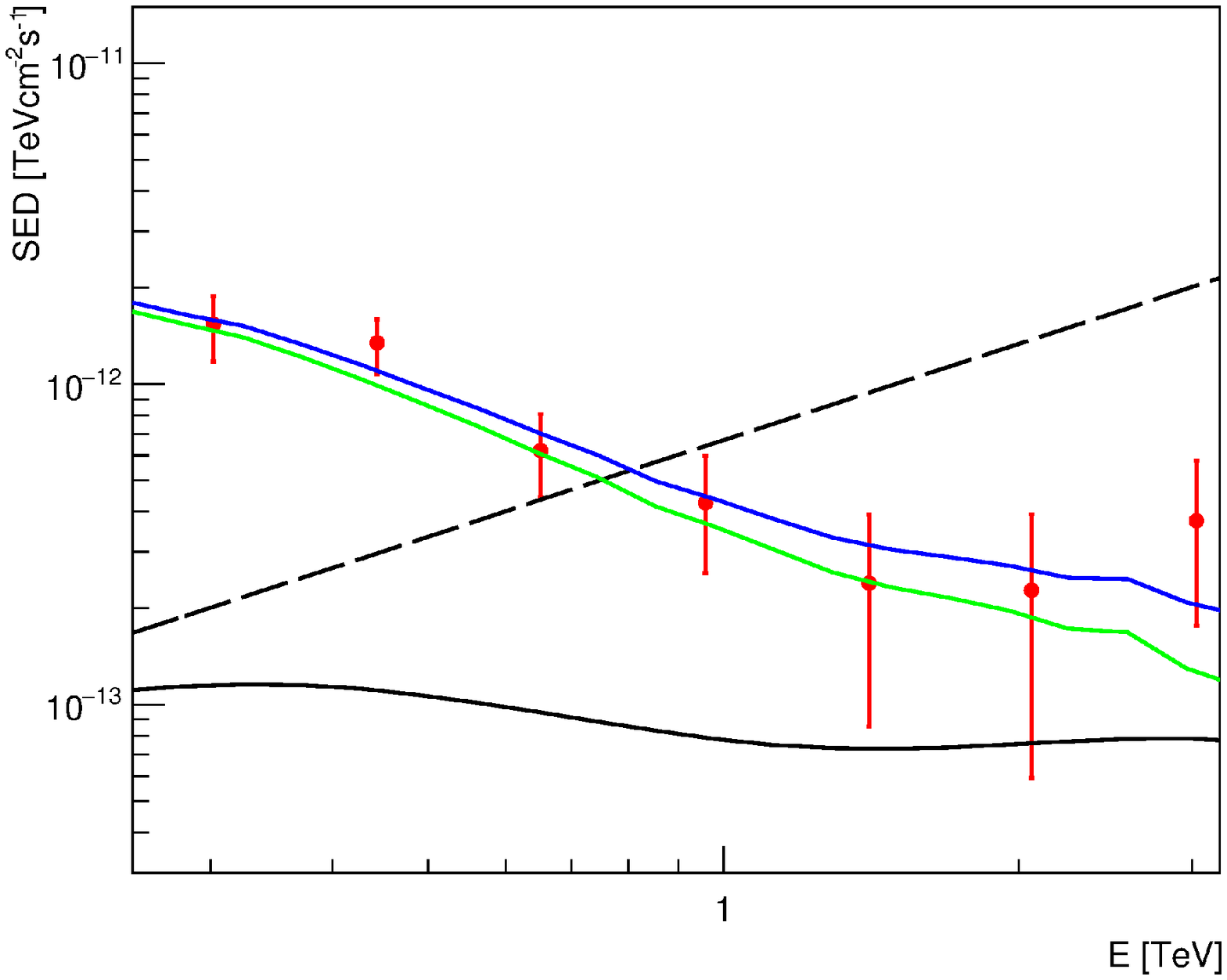}}
\end{minipage}
\hfill
\begin{minipage}{0.50\linewidth}
\centerline{\includegraphics[width=0.9\linewidth]{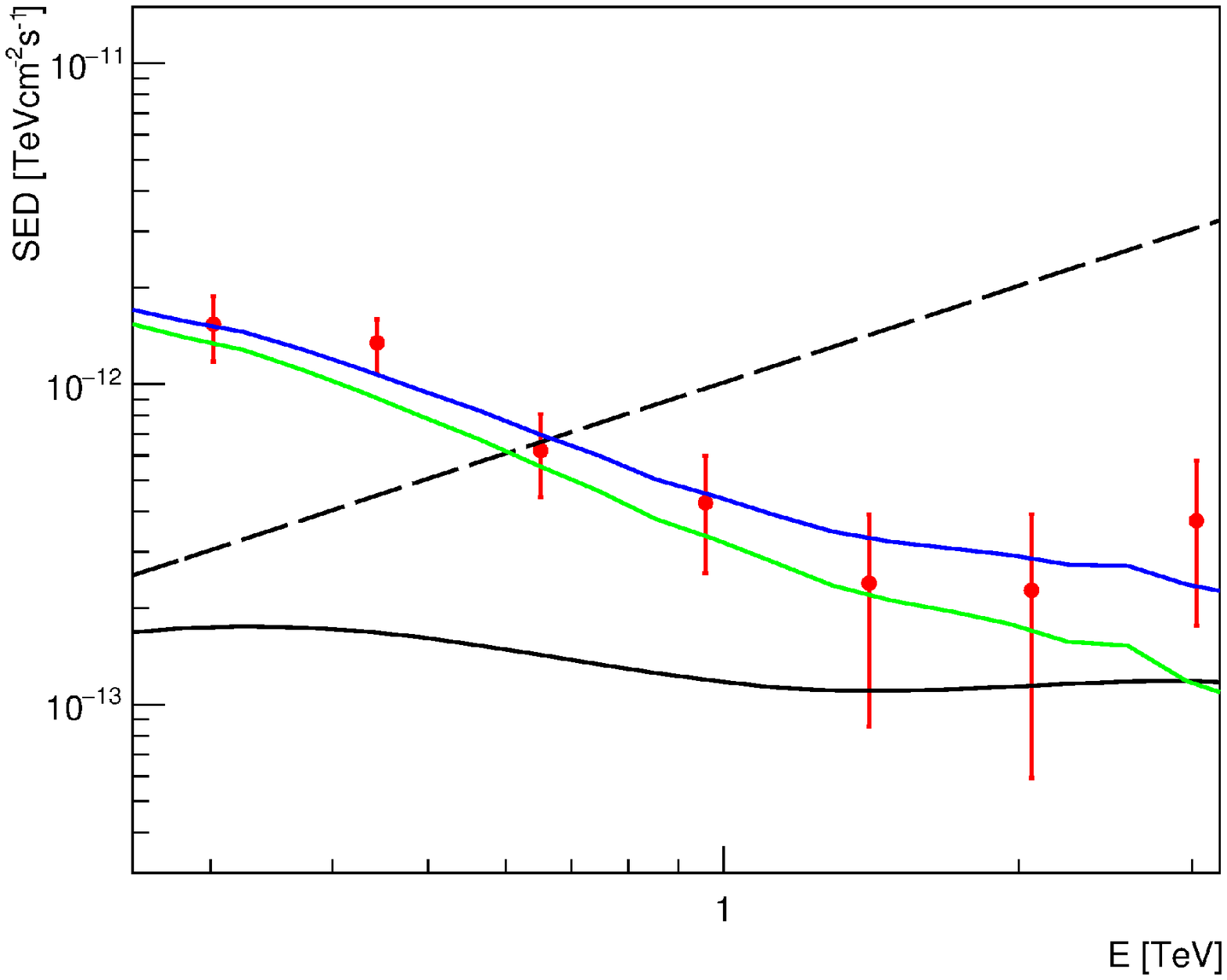}}
\end{minipage}
\newline
\begin{minipage}{0.50\linewidth}
\centerline{\includegraphics[width=0.9\linewidth]{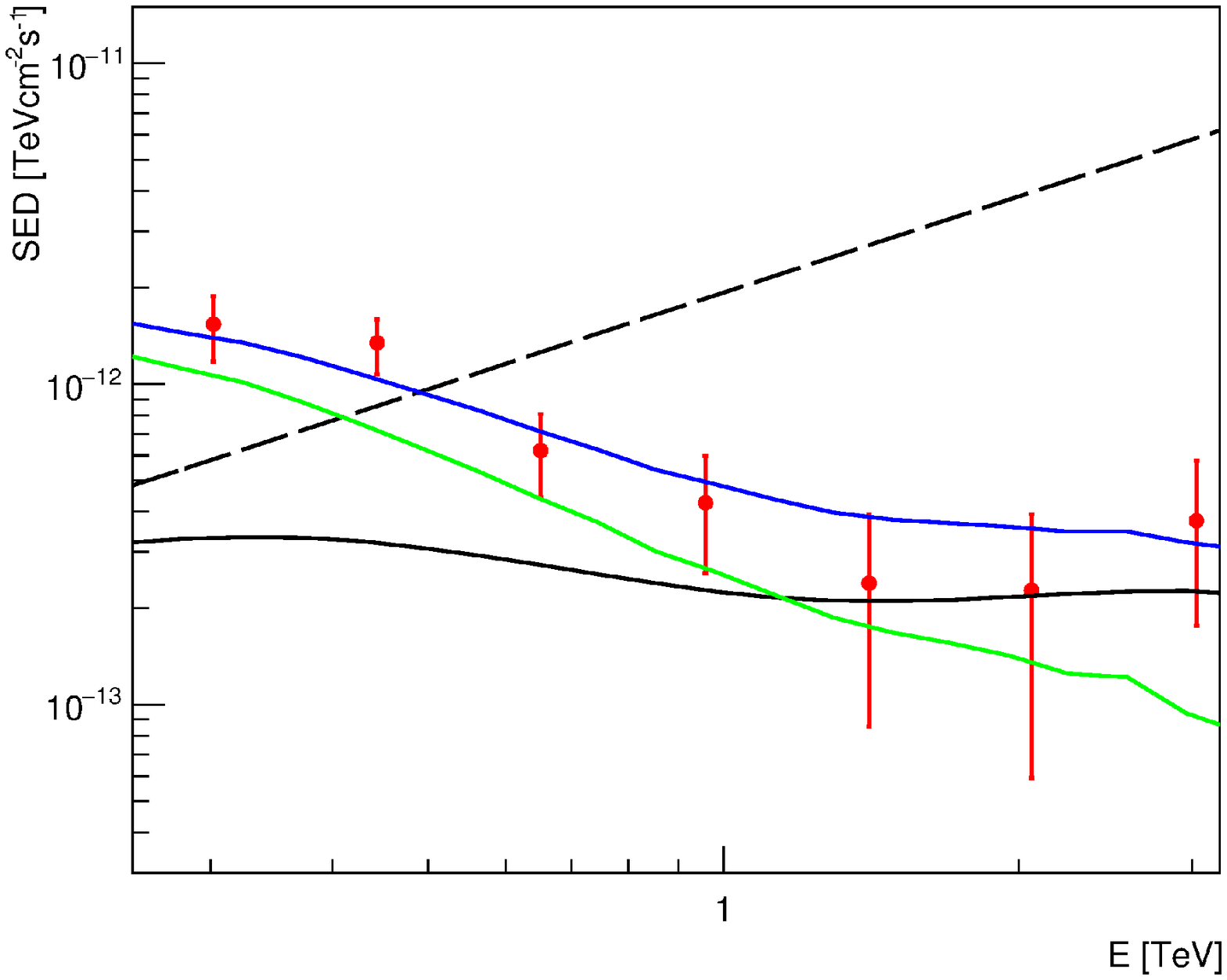}}
\end{minipage}
\hfill
\begin{minipage}{0.50\linewidth}
\centerline{\includegraphics[width=0.9\linewidth]{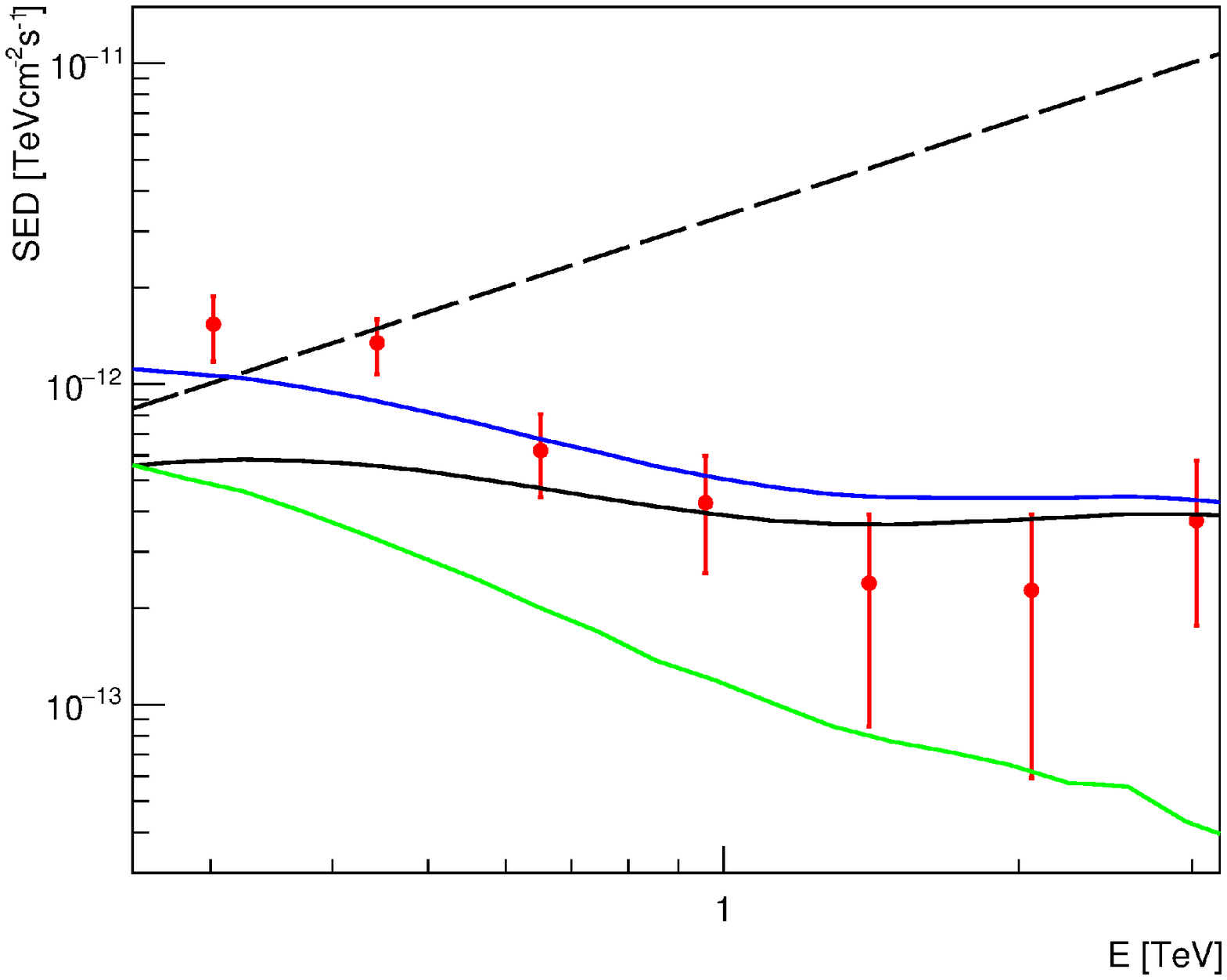}}
\end{minipage}
\caption[]{Fits (solid blue curves) to the observed SED of 1ES 0347-121 (red circles with statistical uncertainties) for $F_{Abs}$=0.3 (top-left), $F_{Abs}$=0.5 (top-right), $F_{Abs}$=0.7 (low-left), and $F_{Abs}$=0.9 (low-right). Intrinsic (primary) spectrum is denoted by dashed black curve, absorbed --- by solid black curve, cascade component --- solid green curve. Calculations were performed with the ECS 1.0 code.}
\label{fig:mult}
\end{figure}

Spectral signatures of blazar emission in the {\it scenario 1} ---namely, 1) a high-energy cutoff, 2) an ``ankle'' formed by the intersection of the primary and cascade components, 3) a possible cutoff of the cascade component caused by the EGMF, and 4) a posible recovery of the primary component at an energy below this latter ``magnetic cutoff'' --- were already considered by us before \cite{r64} (see Fig. 9). In \cite{r38} it was extensively demonstrated that the intersection of the primary and cascade components may account for the anomaly of \cite{r18}, and thus the intergalactic EM cascade development may, to some extent, mimic the $\gamma\rightarrow ALP$ mixing process signature. Here we concentrate on the relative contribution of the primary and cascade components to the observable spectrum in the {\it scenario 2}. Fig.~\ref{fig:mult} shows a model of the observable spectral energy distribution (SED) for blazar 1ES 0347-121 ($z$= 0.188) \cite{r65} and different values of the $F_{Abs}$ parameter defined as the ratio of the primary (absorbed) component to the total observable intensity at the center of the last (high-energy) bin in the observed spectrum of this source. Adiabatic losses are neglected here, as they do not change the shape of the primary spectrum. The contribution of cascades from the low-energy component to the observable spectrum (see Fig. 2) is also neglected. Reasonable fits are obtained for all cases, except the one of $F_{Abs}$= 0.9.

\begin{figure}
\begin{minipage}{0.90\linewidth}
\centerline{\includegraphics[width=0.7\linewidth]{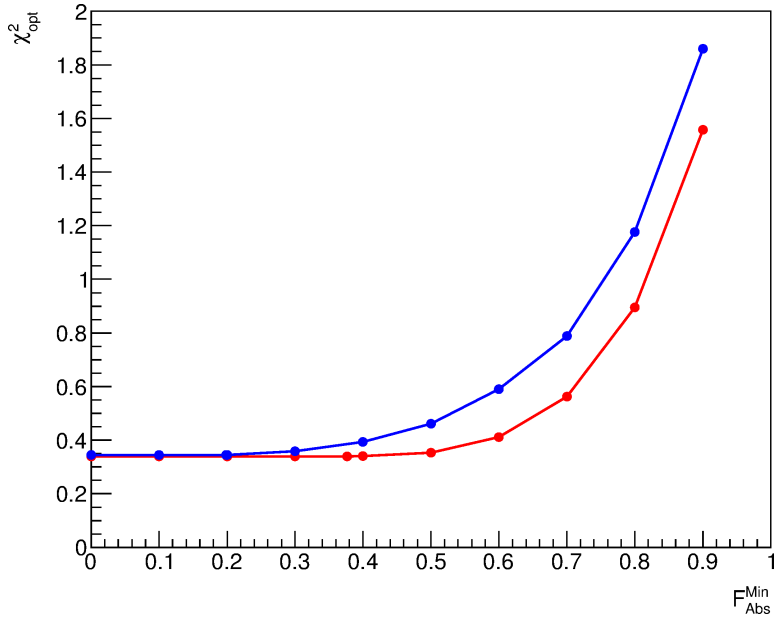}}
\end{minipage}
\caption[]{Dependence of $\chi_{opt}^{2}$ on $F^{Min}_{Abs}$: ECS 1.0 (red circles) and ELMAG 2.02 (blue circles). The lines are drawn merely to guide the eye.}
\label{fig:chisq}
\end{figure}

Fig.~\ref{fig:chisq} shows the dependence of the goodness-of-fit parameter ($\chi_{opt}^{2}$) for 1ES 0347-121 (see Fig.~\ref{fig:mult} for several examples of such fits) on $F^{Min}_{Abs}$, that is, $\chi^{2}$ optimized over the range of $F_{Abs}$ from $F^{Min}_{Abs}$ to 1, obtained with two codes, ECS 1.0 and ELMAG 2.02. Cascade components for both ECS 1.0 and ELMAG 2.02 are from Fig.~\ref{fig:spectra} (cyan circles and histogram, respectively). The actual value of $F_{Abs}$ corresponding to the minimum value $\chi_{opt}^{2}$ is, as a rule, near to the threshold $F^{Min}_{Abs}$. The configurations with high values of $F^{Min}_{Abs}>$0.7--0.8 are disfavoured, indicating that the contribution of the cascade component to the observable spectrum is significant even at $E>$~1~$TeV$. The physical reason for this effect is clear: given the high energy of primary $\gamma$-rays in the {\it scenario 2}, secondary photons are effectively produced even at multi-$TeV$ energies. At the same time, these cascade $\gamma$-rays are absorbed not more strongly than the primary ones (in fact, even slightly weaker, as they travel less distance), therefore their contribution to the observable spectrum may be comparable with that of the primary component even at the highest energy bins accessible to the existing $\gamma$-ray telescopes.

\section{Excess of extreme $TeV$ blazars from the Fermi-LAT distribution on voidiness} \label{sec:flood}

\begin{figure}
\begin{minipage}{0.50\linewidth}
\centerline{\includegraphics[width=0.9\linewidth]{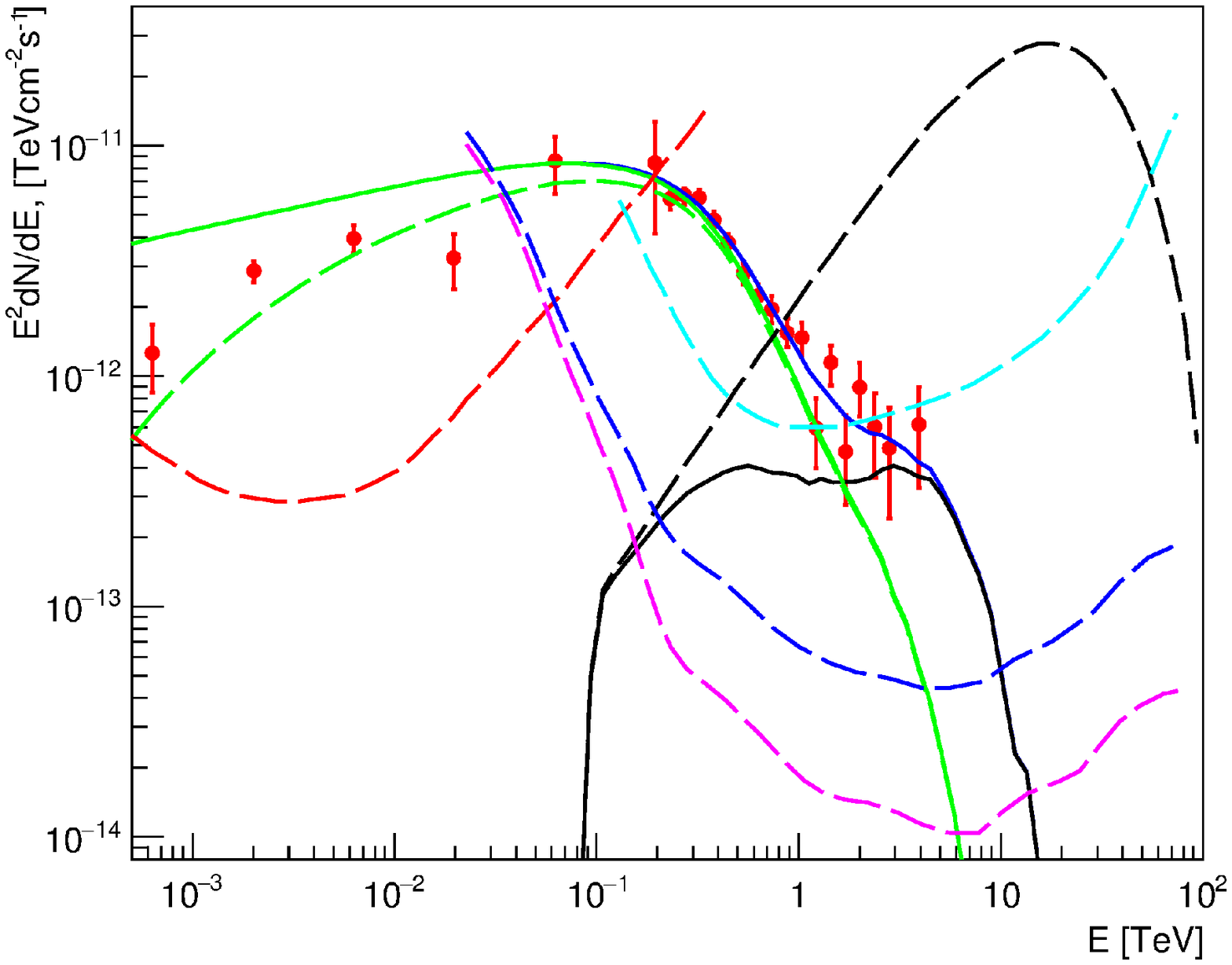}}
\end{minipage}
\hfill
\begin{minipage}{0.50\linewidth}
\centerline{\includegraphics[width=0.9\linewidth]{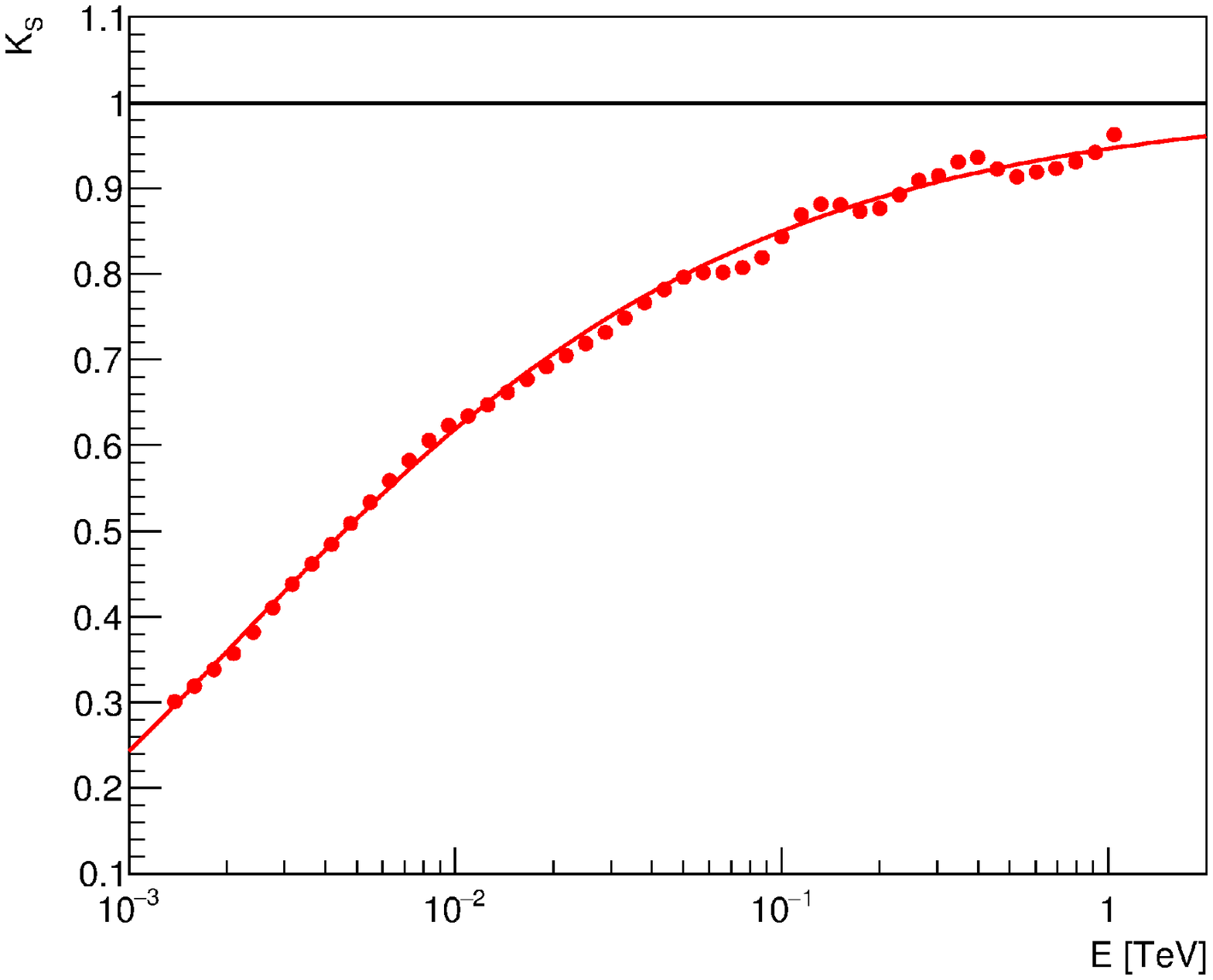}}
\end{minipage}
\caption[]{Left: fits to the observed SED of blazar 1ES 1218+304 obtained with the ELMAG code. The meaning of solid curves and dashed black curve is the same as in Fig.~\ref{fig:mult}. Dashed green line --- cascade component for $B$=$10^{-16}$ $G$, dashed red line --- Fermi-LAT sensitivity (10 years), dashed cyan line --- H.E.S.S. sensitivity (100 hours), dashed blue line --- CTA sensitivity (100 hours), dashed magenta line --- CTA sensitivity (1000 hours). Right: suppression factor for the case of $B$= $10^{-16}$ $G$ (red circles) and its parametrization by a continious function (red curve).}
\label{fig:flood}
\end{figure}

Hard-spectra blazars observed with the Fermi-LAT telescope \cite{r66} tend to be located towards the LSS voids and, moreover, the integral high-energy ($E>$10 $GeV$) flux registered from these active galactic nuclei is typically several times greater for such underdense lines of sight \cite{r25}. This latter strong increase of the observed flux could not be explained by the diminished EBL intensity in voids \cite{r25},\cite{r67}--\cite{r68}. It appears that the effect of \cite{r25} may be explained in the framework of the IECM, assuming that the cascade component dominates the observable intensity at $E\sim$10 $GeV$. Fig.~\ref{fig:flood} (left) shows two fits of observed SED for blazar 1ES 1218+304 (z= 0.182) \cite{r69},\cite{r62} with two options for the cascade component: for $B$=0 (solid green curve) and $B$=$10^{-16}$ $G$ and other parameters according to \cite{r70} (dashed green curve). The suppression factor of the cascade component was estimated using a parametrization of results obtained in \cite{r70} (Fig. 1) with slightly different redshift, $z$=0.14. This parametrization is shown in Fig.~\ref{fig:flood} (right). Sensitivity curves for various instruments and observation times are from \cite{r71} (Fig. 1). Calculations for the case of $B$= 0 are from \cite{r38}. One can see that the appearance of a strong cascade component at low energies could nicely accomodate for the effects found in \cite{r25}. The astrophysical implication of this effect is quite interesting: unless the results of \cite{r25} are caused by a statistical fluctuation, there is a hint for a new blazar population with very hard spectra in the $TeV$ energy region. Some of these sources might be observed with the CTA instrument in future (see dashed magenta curve in Fig.~\ref{fig:flood}). Finally, we note that the part of the cascade component flux between the dashed and solid green curves may create a magnetically broadened pattern around the source, in agreement with results of \cite{r26}. All results presented here and in \cite{r38} do not contradict recent constraints on the EGMF strength and structure (e.g. \cite{r72}, for a recent compilation of results see \cite{r73}).

\section{Conclusions} \label{sec:conclusions}

In this work we have reviewed the main extragalactic $\gamma$-ray propagation models with emphasis on the intergalactic electromagnetic cascade model (IECM). A new Monte Carlo code ECS 1.0 for detailed simulations of observable spectra was developed. Our calculations show that all known deviations from the absorption-only model can be successfully accomodated in the framework of the intergalactic electromagnetic cascade model. Future observations with existing and next-generation instruments such as Fermi-LAT and CTA \cite{r74}--\cite{r75} will allow to confirm or constrain this model.

\section*{Acknowledgments}

This work, for the most part, was performed during the author's stay in the Institute for Cosmic Ray Research, University of Tokyo, Tokyo, Japan. I'm grateful to the members of the ICRR CTA group for extremely helpful discussions, and to the organizers and participants of the Moriond-2017 conference (the VHEPU and EW sessions) for a nice meeting and fruitful discussions. This work was supported by the Students and Researchers Exchange Program in Sciences (STEPS), the Re-Inventing Japan Project, JSPS.

\end{document}